\documentclass[prx,superscriptaddress,twocolumn,nofootinbib]{revtex4-2}

\usepackage{graphicx,epsfig}
\usepackage{color}
\usepackage[usenames,dvipsnames]{xcolor}
\usepackage{amsmath,bbm,amssymb, amsthm}
\usepackage{dsfont} 
\usepackage{stmaryrd}
\usepackage{comment}
\usepackage{hyperref}
\usepackage{changepage}
\usepackage{numprint}

\makeatletter
\newcommand{\fmarki}{\ensuremath{\dagger}}
\newcommand{\fmarkii}{\ensuremath{\ddagger}}
                
\def\@fnsymbol#1{{\ifcase#1\or \fmarki\or \fmarkii \else\@ctrerr\fi}}
\makeatother

\def\bb{\begin{eqnarray}}
\def\ee{\end{eqnarray}}

\begin{document}

\title{Reply to ``Comment on `Extending the laws of thermodynamics for arbitrary autonomous quantum
systems'''}

\author{Cyril Elouard$^{*}\,$}
\email{cyril.elouard@gmail.com}

\affiliation{LPCT, CNRS, Université de Lorraine, 54506 Vand{\oe}uvre l\`es Nancy, France}

\author{ Camille Lombard Latune$^{*}\,$ }
\email{cami-address@hotmail.com}
\thanks{\\$^{*}$ These two authors contributed equally.}
\affiliation{ENSL, CNRS, Laboratoire de physique, F-69342 Lyon, France}

\begin{abstract}
    In his Comment [1], Philip Strasberg (PS) argues from the analysis of different examples that the framework we have presented in [2] does not recover known results of macroscopic textbook thermodynamics. Here, we show that such apparent contradictions disappear when the necessary assumptions the aforementioned known results pre-suppose are applied. Those assumptions concern the control ability of the observer, the nature of the described degree of freedom, or the scale of the systems. The ability to relax those assumptions is precisely a motivation of our framework, which can explore the capacity of quantum systems to exchange work and heat even at scales not captured by textbook thermodynamics. We take the opportunity of this reply to further expand on the use of our framework and its connections with traditional thermodynamics.
\end{abstract}

\maketitle

In his Comment [1], Philip Strasberg (abbreviated PS in the following) argues that the framework we have presented in [2] is “in conflict with textbook thermodynamics”. He provides several examples for which he claims that our framework does not yield the physically expected behavior, or provides a wrong estimate of entropy production.

We show below that his conclusions are obtained by incorrectly comparing our findings with intuitions from classical macroscopic thermodynamics without applying the necessary assumptions they pre-suppose. 
More precisely, the framework we introduced [2] provides the flexibility to describe all the resources that can be stored in a quantum system and whose consumption is equivalent to work. One of the novel possibilities opened by our framework is the analysis of completely autonomous quantum machines, composed of several systems which can be efficiently manipulated locally. This is therefore a natural emphasis of our article [2]. 

Nevertheless, as we mention in [2], our results hold in principle for larger scale systems, up to the scales at which thermodynamics has initially been developed. However, when increasing the system size, it becomes natural to expect that only partial control is practically possible (meaning that only certain types of nonequilibrium resources can be manipulated in practice). In addition, still with increasing system size, new phenomena such as equilibration of an isolated system become possible, when the relevant degrees of freedom are described. Finally, again in a perspective of very large systems, new scaling properties emerge, e.g. coupling energies become typically negligible in front of bulk. All these important assumptions connect microscopic quantum mechanical description to phenomena of the macroscopic world. As our approach is, by design, formulated at the quantum-mechanical level, it is natural that these assumptions must be added on top of our framework to address these macro-scale phenomena. This task was beyond the scope of our first article [2], except for the notion of partial control for which we provided a methodology in Appendix D.\\  

In the remainder of this reply, we analyze the examples mentioned by PS to show that such assumptions can be added to our formalism to describe those macroscopic situation if needed. Conversely, our framework allows to selectively relax those assumptions of textbook thermodynamics, allowing to analyze new behavior relevant at the quantum scale.

For pedagogical purposes, we start with the examples of two systems in a pure state, which was mentioned by PS as a criticism of the notion of effective temperature we use.

\subsection*{Ex1: Two systems initially in their ground state} 

As PS points out in [1], traditional thermodynamics predicts that two identical systems initially in their ground state (or equivalently, in thermal equilibrium states at equal vanishing temperatures) should exchange no heat flow when they are put in contact. We emphasize that this statement is derived for two macroscopic systems (in the thermodynamic limit).

In contrast, our framework allows us to analyze the case where we couple two quantum systems, whatever their size. Starting from the two systems each in the ground state of its Hamiltonian, we consider that a coupling Hamiltonian is switched on at time t=0. If the coupling term does not commute with the two local Hamiltonians (the only non-trivial case), the two systems are at $t=0^+$ in an out-of-equilibrium state, which starts evolving for $t>0$: in general, the systems’ energies and entropies will vary, and as pointed out by PS, there will be an increase of the systems' effective temperatures between time $t=0$ and $t>0$, which we interpret as heat flowing into those two systems. Energy conservation in the total system implies that the increase of the system’s local energies is compensated by a decrease of the interaction energy. 
Does the conversion of coupling energy into heat provided to the two systems contradicts textbook thermodynamics? Our claim is that it does not violate the second law as it can be seen from Eq.~(11) of [2] which shows that consumption of coupling energy is equivalent to work expenditure, and it can therefore be transformed into heat even though the two systems are both at the same temperature. \\

Now, how can we reconcile this view with traditional macroscopic thermodynamics?
To do so, we have to consider the same limit of two macroscopic systems, such that the coupling energy-scale is negligible with respect to their internal energies. In this limit, the variation of energy and therefore of temperature during the evolution is essentially zero, hence a zero heat exchange. This is also the limit in which the zeroth law is satisfied. Rather than being in disagreement with textbook thermodynamics, Eq.~(5) of [2] provides a quantitative description of the departure from this textbook situation when the size of the considered systems is finite.

\subsection*{Ex2: Two systems initially in a pure state}
We now extend the analysis to the case of two systems initially in pure states of different energies, also mentioned by PS in [1]. PS claims: “Then, the flow of energy is determined by their energy difference, which is not predicted by the ELL temperature”. We disagree with this analysis and emphasize that the difference in the effective temperatures of the two systems determines the direction of the flow of \emph{heat}, not energy. There is no reason to assume that the two systems will exchange only heat. In contrast, their out-of-equilibrium initial state let us expect that they can provide work. The latter can then be consumed to induce a heat flow which does not necessary follow the initial temperature bias (in our formalism, both systems have initially zero effective temperature if they are in a pure state). Our framework is precisely designed to account for all these situations, not only the case where the two systems exchange only heat.

Nonetheless, to show the flexibility of our framework, we explain how the case of a pure exchange of heat can be retrieved. PS does not provide in is comment any precise context in which a pure state can be treated as a source of heat. This occurs, to our knowledge, only in the case of a macroscopic interacting many-body systems which is ``observed locally".  More precisely, unitary thermalization has been demonstrated in the specific sense that the stationary statistics of a local observable of the system would match canonical thermal equilibrium distribution at a temperature determined by the extended system’s average energy. We then assume that the “empirically measured temperatures” mentioned by PS refer to the temperature determining this effective statistics, which can be accessed via local observable measurements.

	Then, when considering two such equilibrated many-body systems $S_A$ and $S_B$ and coupling them via a local interaction, one can expect at the interface between the two systems a heat flow determined by the local-temperature bias. We see that here textbook thermodynamic principles are valid for a specific choice of systems of interest, namely, local degrees of freedom. This is the necessary price to define thermal equilibrium in a closed unitary setup.
	In perfect agreement, this situation would be efficiently captured in our formalism by considering subparts $A$ and $B$ (of $S_A$ and $S_B$) corresponding to the same local degrees of freedom involved in the local coupling and obeying canonical statistics.
 More precisely, one could use Eq.(15) of [2], accounting for heat transfers between A and B, but also between A and the rest of $S_A$, and between B and the rest of $S_B$). 

\subsection*{State of knowledge and control ability}

The need to properly choose the system’s boundary to obtain the expected thermodynamic behavior might appear, at first glance, to be a limitation of our formalism. Actually, it reflects a universal necessary feature of thermodynamics: The expression of entropy production depends on the state of knowledge and the control ability of the observer, as PS points out in his comment.
In the above example (Ex2), it is expected that the information about the nonequilibrium-ness of the state of $S_A$ and $S_B$ quickly becomes inaccessible if one only have local access/control and hence an impossibility to extract work. In contrast, a non-local coupling, or a local coupling between two smaller quantum systems could give access to the resources contained in those nonequilibrium states (quantified by the generalized ergotropy discussed in Section V of [2]), leading to a behavior that does not correspond to a spontaneous heat flow determined by only the initial energy-bias.

More generally, Eqs.(5)-(11)-(18) of our article are taking into account all the resources that can in principle be accessed in each of the systems, hence keeping only the most fundamental source of irreversibility. This is, by design, the opposite limit to that PS is considering in several examples of his comment [1]. Our motivation was precisely to quantify those resources that become relevant when one departs from the case of a macroscopic system controllable only via classical means. For pedagogy, we insisted in most of the article [2] on the case where all the resources are \emph{a priori} considered to be accessible, in order to highlight the novelty of our standpoint. We acknowledge that this limit corresponds in practice to systems of limited size and complexity. Nonetheless, we explain in Appendix D a methodology to interpolate within our framework between the full-control situation and the no-control case, by progressively incorporating resources that are within the reach of a given setup (e.g. the coherent displacement of a bosonic mode) while ignoring the others. Once again, our approach constitutes an extension of the usual thermodynamic settings, that departs \emph{on purpose} from the fully macroscopic limit. This said, we acknowledge the practical interest of investigating more methods to implement the coarse-grainings reflecting limited control ability on complex quantum systems. Developing such methods, or combining our framework with the methods proposed in the literature (in particular the one proposed by PS and cited in his comment [1]) is, in our minds, an exciting follow-up of [2]. Actually, as noticed by PS, the entropy-based definition of effective temperature can easily be extended to coarse-grained choices of entropy as the observational entropy, opening perspectives to combine the generalized notion of work authorized by our framework with a simplified description of complex quantum systems.

To address more precisely PS’ criticisms about the level of control, we now analyze the case of the expansion of a gas.

\subsection*{Ex3: expansion of a gas}

When considering the expansion of a macroscopic gas B in contact with a small system A, the origin of the extensive entropy production mentioned by PS is the internal dynamics of B, and the role of the coupling to A is essentially negligible.
The same question could be then be asked by considering system B alone, evolving freely (unitarily) from an out-of-equilibrium initial state. In this situation, Eqs.~(5) or (18) of [2] predict no entropy production at all, because in principle all the resource initially present is conserved. 
Of course, when increasing the complexity of B and the atypicallity of its initial state, one can expect that the degree of control necessary to extract this resource will become unrealistic. 
In this situation, we agree that some constraint must be added to take into account the finite degree of control of the observer. The observational entropy proposed in [3] is a possible approach. In the appendix D, we propose another method that is directly fitted to be used within our formalism. For instance, one could imagine taking advantage of bias between local pressure into the “gas” if its equilibration is slower than our ability to control it. Crucially, this methodology allows us to consider all situations where part of the resource stored in $B$ remains accessible, by introducing unitary transformations $U[\alpha]$ which correspond to the control operations that are available to the observer. In the extreme limit of a system equilibrating immediately, the initial resources becomes immediately unavailable to the observer. This means that the optimal unitary $U[\alpha]$ extracting the largest amount of available resource from the system B quickly becomes close to the identity under the free dynamics of B. When this is the case, our formulation of the second law becomes the Eq.(1) of [2] which provides an extensive scaling of entropy produced in systems treated as heat reservoirs, as analyzed in the Ref.~[4] (cited as Ref.[3] in [2]). 

To conclude, rather than being “in contradiction” with this notion of extensive entropy production occurring in many situations of thermodynamics, our framework enables a powerful and flexible analysis of new situations where increased control of the system $B$ (or where its finite-ness) can yield departure from this scaling. The Eq.(18) of [2], used by PS to formulate his criticism, correspond to the limit of full control, 
where entropy production is only due to correlation buildup between systems.

\subsection*{Other criticisms}
We finish by replying to other technical criticisms raised by PS. 
\subsubsection*{1) Spontaneous entropy increase}
PS points out that “by time reversal symmetry violations of Eq. (1) [of Ref.~[1] ] must be possible, and all what one can hope for is that an increase in Eq.~(1) is much more likely than a decrease for arbitrary states. Already this basic insight casts doubts on the validity of the ELL framework, which predicts that a spontaneous decrease in Clausius’ expression is impossible (and not just extremely unlikely) for all de-correlated states”. 
In contrast, we emphasize that the formalism we exposed in Ref.~[2] is meant to capture the average thermodynamics, not single fluctuation. Eqs.(5)-(11)-(18) then quantify the entropy production averaged over realizations. Therefore, no violation of these inequalities are expected.
Contrary to PS’ claims, application of time-reversal to a case where entropy production is strictly positive according to Eq.~(18) of [2] (i.e. to Eq.~(1) of [1]) does not yield any violation of the Second law: Indeed, in this case, the initial state of the time-reversed trajectory must necessarily be correlated, a case which is excluded by the assumptions leading to Eq.~(18) of [2]. Instead, Eq. (17) of [2] can be applied to describe the time-reversed trajectory.

\subsubsection*{2) Zero effective temperature limit and degenerate ground state}

PS mentions that Eqs.~(5)-(18) diverge in the limit of vanishing effective initial temperature $\beta_B(0)\to\infty$. In addition, he points out that the definition of effective temperature, as formulated in the manuscript, may not always yield a well-defined temperature in the case of a Hamiltonian with degenerate ground state. We show here than simple reformulations of Eqs.~(5)-(18) and a slight extension of the definition of the effective temperature enable to include those cases, without affecting the discussions present in [2].\\

As a first step, we make PS's argument more explicit by noting that the definition of effective temperature breaks down in the case where there is no choice of inverse temperature $\beta_B$ such that $S[w_B[\beta_B]]=S_B(t)$. Denoting $d_g$ the ground state degeneracy of system $B$ (i.e. dimension of the eigenspace of lower energy), then the entropy of the thermal state $S[w_B[\beta_B]]$ is lower-bounded by $\log d_g$, and this bound is saturated in the limit $\beta_B\to \infty$.\\

Below, we show that by reformulating Eq.~(5) and (11) and then taking the vanishing temperature limit, one obtains the following inequalities, that we interpret as expressions of the second law in the zero temperature limit:
\bb \label{eq:0}
Q_B &\leq &  0\nonumber\\
W_B  &\geq & \Delta E_A + \Delta E_\text{int}.
\ee
In addition, by re-defining the effective temperature $T_B(t)=\beta^{-1}_B(t)$ as the inverse of the positive number $\beta_B(t)>0$ minimizing the difference $|S[w_B[\beta_B(t)]]-S_B(t)|$ (rather than requiring the equality), we find that the same equations describe the case of a system $B$ with degenerate ground state $d_g>1$ and initial entropy below $\log d_g$.

With this modified prescription for the effective temperature, we see that two cases are possible. Case 1: $S_B(t) \geq \log d_g$ (this was the case implicitly considered in [2]). In this case there is a unique solution $\beta_B(t)$ to the equation $S_B=S[w_B[\beta_B(t)]]$. The solution may be $T_B(t)=0$ if $S_B(t)=0$ and $d_g=1$. Case 2: $S_B(t) < \log d_g$, which requires $d_g>1$. Then $T_B(t) = 0$ and $S_B < S[w_B[\beta_B(t)]]=\log d_g$. 
To capture both situations, we introduce $\zeta_B(t) = S[w_B[\beta_B(t)]]-S_B(t)$, which fulfills:
\bb
\zeta_B(t)  = \left\{\begin{array}{ll}
   0 & \text{when $S_B(t) \geq \log d_g$}   \\
   \log d_g-S_B(t)   > 0  & \text{when $S_B(t) < \log d_g$}
\end{array}\right.
\ee

We now generalize Eq.(5) and (11) of [2] in the zero-temperature limit. We first consider a strictly positive effective temperature $T_B(0)$ and multiply Eq.~(7) of [2] (which simply encodes the facts that systems $A$ and $B$ are initially un-correlated and evolve unitarily), to obtiain:
\bb
T_B(0)\Delta S_A = T_B(0) I_{AB}(t) - T_B(0)\Delta S_B(t).\label{eq1}
\ee
Then, we use our definition of $\zeta_B(t)$ to show:
\bb 
\Delta S_B = \Delta S[w_B[\beta_B(t)]]-\Delta \zeta_B.
\ee
Keeping the same definition of $E_B^\text{th}(t)=\text{Tr}\{H_B w_B[\beta_B(t)]\}$ and following the steps of the derivation of Eq.(5) in [2]
\begin{widetext}
\bb
T_B(0)\Delta S_A + \Delta E_B^\text{th} - T_B(0)\Delta \zeta_B(t)  = T_B(0) I_{AB}(t) +  T_B(0) D(w_B[\beta_B(t)]\|w_B[\beta_B(0)]) \geq 0\label{eq:ZeroTEP2}
\ee
\end{widetext}

We now note that even though the relative entropy $D(w_B[\beta_B(t)]\|w_B[\beta_B(0)])$ diverges in the limit of zero initial effective temperature $\beta_B(0)\to\infty$, we have $\underset{T_B(0)\to 0 }{\lim} T_B(0) D(w_B[\beta_B(t)]\|w_B[\beta_B(0)]) = E_B^\text{th}(t)$, which is finite. In addition, as $\vert \zeta_B\vert$ is upper-bounded by $\log d_g$, we have $\underset{T_B(0)\to 0 }{\lim} T_B(0)\Delta \zeta_B(t) = 0$. Finally, Eq.~\eqref{eq:ZeroTEP2} admits a well defined limit for $T_B(0)\to 0$ which verifies:
\bb 
Q_B(t)  \leq 0, \quad\forall d_g\in\mathbb{N}.\label{eq:QBneg}
\ee
We see that the case of a degenerate ground state does not affect the existence of the zero-temperature limit, and the situation where the effective temperature was ill-defined according to the prescription of [2] can be described by setting the effective temperature to $0$.

Eventually, we combining $\Delta E_A + \Delta E_B + \Delta E_\text{int} = 0 $, $\Delta E_B = - Q_B(t) - W_B(t)$, and Eq.~\eqref{eq:QBneg} to deduce 
 the inequality $W_B  \geq \Delta E_A + \Delta E_\text{int}$, which is the low-temperature limit of Eq.(11) of [2].\\

\subsubsection*{3) Temperature from the partial derivative of the entropy}

Finally, we refute PS's argument that our definition of effective temperature violates the connection between temperature and a partial derivative of entropy with respect to energy. We start by recalling that the equality mentioned by PS is a consequence of the thermodynamic identity, which for a classical system reads $dE = TdS - PdV$, and is valid for an infinitesimal transformation along which temperature and pressure remain well-defined for the system. Then, the temperature can be defined as the partial derivative of the energy $E$ with respect to entropy, \emph{at constant volume and particle number, i.e. in the absence of work exchange}, or equivalently $1/T = \frac{\partial S}{\partial E}\vert_{V,N}$. Our definition of temperature directly fulfills this definition since, in the absence of work, it is always true that $dE_B = T_BdS_B$ (see Eq.~(9) of [2]). More generally, we can write down an equivalent of the thermodynamic identity (for system $B$):
\begin{equation}
dE_B = T_BdS_B - \delta W_B. 
\end{equation}
The first term corresponds to the variation of the thermal energy of $B$, and therefore matches the infinitesimal heat. The second term verifies $-\delta W_B = d(E_B-E_B^\text{th})$, which is the variation of the nonequilibrium resources and corresponds to work exchanges (a generalized version of terms like $-PdV$ or $\mu dN$ which assume specific mechanisms for work exchanges). Approaching zero temperature (or pure state) does not invalidate this identity. \\

We finish this reply by apologizing if the results of [3] are incorrectly captured by the last paragraph before the section IX of [2]. We stress however that, when introducing Eq.~(53) in [3] (which has the same form as Eq.(18) of [2]), the authors explicitly mention a restriction to a subclass of initial states for the systems given by Eq.~(44) of [3]. This class is then extended in later Section VII to states containing correlations between A and B, but where B must be diagonal in the energy eigenbasis (see Eq.~(66) of [3]), at the price of additional terms in the expression of entropy production (Eq.~(67) of [3]).
We also acknowledge that we failed to cite [Fannes el al. 2013] when listing previous use of the same definition of effective temperature, even though this article is part of the list of references cited in [2] under label Ref.~[43]. \\

In conclusion, we have shown that the framework we introduced in [2] indeed leads to expected textbook thermodynamic results in the examples mentioned by PS in [1], provided the assumptions connecting quantum mechanics to the macroscale are also applied. In contrast, our approach allows to relax part or all of these assumptions and therefore describe situations beyond the scope of textbook thermodynamics, which is precisely our motivation. We also showed that the special limit of vanishing initial effective temperature can be taken if needed. We acknowledge that taking into account the possibility of a degenerate ground state requires a slight modification of the definition of the effective temperature, which, as we showed, has no practical consequences whatsoever.

\vspace{2cm}

References:\\

[1]  Philipp Strasberg, "Comment on "Extending the Laws of Thermodynamics for Arbitrary Autonomous Quantum Systems"", \href{https://doi.org/10.48550/arXiv.2309.04170}{arXiv:2309.04170}.\\
 
[2]  C. Elouard and C. Lombard Latune, “Extending the Laws of Thermodynamics for Arbitrary Autonomous Quantum Systems,” \href{https://journals.aps.org/prxquantum/abstract/10.1103/PRXQuantum.4.020309}{PRX Quantum 4, 020309 (2023)}.\\

[3] P. Strasberg and A. Winter, “First and Second Law of Quantum Thermodynamics: A Consistent Derivation Based on a Microscopic Definition of Entropy,” \href{https://journals.aps.org/prxquantum/abstract/10.1103/PRXQuantum.2.030202}{PRX Quantum \textbf{2}, 030202 (2021)}.\\

[4] K. Ptaszy\'nski and  M. Esposito, ``Entropy prodution in open systems: The predominent role of intraenvironment correlations'', \href{https://journals.aps.org/prl/abstract/10.1103/PhysRevLett.123.200603}{Phys. Rev. Lett. \textbf{123}, 200603 (2019)},

\end{document}